\documentclass{aa}
\usepackage{natbib}
\usepackage{psfig,graphicx,epsfig}

\def\gsim{\;\lower4pt\hbox{${\buildrel\displaystyle >\over\sim}$}\,}
\def\lsim{\;\lower4pt\hbox{${\buildrel\displaystyle <\over\sim}$}\,}

\def \xmm {{\em XMM-Newton}}

\def \hcm {\hbox {\ifmmode $ atom cm$^{-2}\else atom cm$^{-2}$\fi}}

\def \arcsec {\hbox{$^{\prime\prime}$}}
\def \degrees {\hbox{$^{\circ}$}}

\def\approxgt{\mathrel{\hbox{\rlap{\lower.55ex \hbox {$\sim$}}
        \kern-.3em \raise.4ex \hbox{$>$}}}}
\def\approxlt{\mathrel{\hbox{\rlap{\lower.55ex \hbox {$\sim$}}
        \kern-.3em \raise.4ex \hbox{$<$}}}}

\newcommand\U[1]{{\,\rm #1}}

\newcommand\rs[1]{_\mathrm{#1}}
\newcommand\tauone{\tau\rs{1\,keV}}

\newcommand\tht{\theta}

\newcommand\amax{a\rs{max}}

\newcommand\thtsc{\theta\rs{scal}}

\def\lsim{\;\raise0.3ex\hbox{$<$\kern-0.75em\raise-1.1ex\hbox{$\sim$}}\;}
\def\gsim{\;\raise0.3ex\hbox{$>$\kern-0.75em\raise-1.1ex\hbox{$\sim$}}\;}

\def\beq{\begin{equation}}
\def\enq{\end{equation}}
\def\begar{\begin{eqnarray}}
\def\endar{\end{eqnarray}}
\def\mathnew{\mathsurround=0pt}
\def\simov#1#2{\lower .5pt\vbox{\baselineskip0pt \lineskip-.5pt
        \ialign{$\mathnew#1\hfil##\hfil$\crcr#2\crcr\sim\crcr}}}

\def \chan {{\it Chandra}}
\def \xmm {{\it XMM-Newton}}
\def \src {G21.5--0.9}

\begin{document}

%\thesaurus{(02.01.1; 02.18.5; 09.03.1; 09.03.2; 09.09.1: \src; 09.19.2)}

\title{The nature of the X-ray halo of the plerion G21.5-0.9 unveiled by XMM-Newton and Chandra}

%\\ {\bf  Draft, 9/7/2001 } \\
%\end{center}

\author{F. Bocchino\inst{1}
\and E. van der Swaluw\inst{2,3}
\and R. Chevalier\inst{4}
\and R. Bandiera\inst{5}
}

\institute{
       INAF-Osservatorio Astronomico di Palermo, Piazza del Parlamento 1,
       90134 Palermo, Italy
\and 
       FOM-Institute for Plasma Physics Rijnhuizen, 
       PO Box 1207, 3430 BE Nieuwegein, The Netherlands
\and 
       School of Physics and Astronomy, The University of Leeds, 
       Woodhouse Lane, Leeds LS2 9JT, UK
\and 
       Department of Astronomy, University of Virginia, 
       P.O. Box 3818, Charlottesville, VA 22903, USA
\and 
       INAF-Osservatorio Astrofisico di Arcetri, Largo E. Fermi 5,
       50125 Firenze, Italy
}

\offprints{e-mail: bocchino$@$astropa.unipa.it}

\date{Received 14 February 2005 / Accepted 15 July 2005}

%\maintitlerunninghead{}
\authorrunning{Bocchino et al.}
\titlerunning{X-ray halo of \src}

\abstract{The nature of the radio-quiet X-ray halo around the plerionic
SNR \src\  is under debate. On the basis of spatial and spectral analysis
of a large \chan\  and \xmm\  dataset of this source, we have developed a
self-consistent scenario which explains all the observational features. We
found that the halo is composed by diffuse extended emission due
to dust scattering of X-rays from the plerion, by a bright limb which
traces particle acceleration in the fast forward shock of the remnant,
and by a bright spot (the ``North Spur'') which may be a knot of ejecta in
adiabatic expansion. By applying a model of interaction between the PWN,
the SNR and supernova environment, we argue that \src\  progenitor may
be of Type IIP or Ib/Ic, and that the remnant may be young (200--1000 yr).

}

\maketitle

\section{Introduction}

The plerionic supernova remnant (SNR) \src\  has been extensively
studied in radio (see e.g. \citealt{bk76,bs81,fhm88,kas92,bwd01}
and references therein) and in the X-ray band
(e.g. \citealt{dsb86,ak90,scs00,wbb01,shp01}). In
spite of several efforts, its pulsar remains undetected
(\citealt{bl96,lm02}). \citet{wsp97} include it among the non-Crab like
class of plerions, because of its low frequency spectral break, for which a non
standard evolutionary path of the pulsar output must be invoked. However,
\citet{bwd01} and \citet{bnc01} pointed out that new observations,
at 94 and 230 GHz respectively, suggest a spectral break above 100 GHz,
much higher than previously thought.

A set of new and detailed observations in the X-ray band performed
with \xmm\  and \chan\  have apparently raised new and interesting
questions about the nature of this object, and in particular on
the extended and diffuse X-ray halo which seems to surround this
plerion. \citet{scs00} seem to have been the first to detect X-ray
emission extending beyond the boundary of the radio plerion. The short
\chan\  calibration observation they used prevented a detailed study
of the halo, and it was not recognized if the emission was thermal or
non-thermal. Due to the lack of further data, \citet{scs00} suggested
that the halo may represent the shell formed by the interaction of the
main blast wave with the surrounding medium. They examined archival
VLA radio data and concluded that the upper limit to the 1 GHz surface
brightness ($1~\sigma$) is $4\times 10^{-21}$ W m$^{-2}$ Hz$^{-1}$
sr$^{-1}$. \citet{wbb01} used the \xmm\  calibration observation of \src\
we also use, and established the non-thermal nature of the X-ray emission
of the halo. The lack of line emission  in the integrated spectra of the
halo pointed toward a very small ionization time ($\sim 3\times 10^8$
cm$^{-3}$ s), if non-equilibrium of ionization were used in the fit.
\citet{wbb01} also noted that the size of the X-ray halo exceeds that
of the radio PWN (pulsar wind nebula) by a factor of 4, a feature which
is not expected and indeed not observed in any other PWN.
They also detected a bright spot (named ``North Spur'') and
some filaments in the halo. \citet{shp01} confirmed the \citet{wbb01}
findings and considered the halo as an extension of the plerion nebula,
but they also pointed out that the observed morphology is puzzling and
cannot be explained by diffusion models. The X-ray morphology was
also briefly discussed by \citet{bwd01}, who confirmed the previous
upper limit on $\Sigma\rs{1~GHz}$, and who suggested multiple events
or injection epochs, noting, however, that the new radio data do not
address the question of why the morphology is so peculiar.  \citet{bb04}
proposed that the X-ray halo is due to photons scattered by ISM dust along
the line of sight, and showed that this model is broadly compatible with
the observed absorption and radial surface brightness profiles. However,
their fit to the profiles showed residuals due the presence of the
(then undetected) weak features in the halo, and an error in the profile
normalization led to the conclusion that an intrinsic halo was required
to fit the data in the 5--8 keV.
Finally,
\citet{b05} has reported the detection of a weak thermal component in
the X-ray spectrum of the ``North Spur'' as seen by both \chan\  and
\xmm. While the metal abundances seem to indicate that the North Spur
is an ejecta knot, the measured X-ray temperature is much lower than
the expected ejecta temperature behind a reverse shock.

It is clear that the nature of the X-ray halo of \src\  is still poorly
known. Most of the studies have been focused on the central plerion and
there has been no systematic study of the halo itself.  As a consequence,
the three possible models for the X-ray halo introduced above, namely the
non-thermal shell, the extension of the plerion and the dust-scattering
of X-rays from the center, have not been properly investigated.  In this
paper, using new X-ray datasets that have been accumulated by \chan\
and \xmm, we propose a possible scenario which explains the halo as
a non-thermal shell superposed on a dust-scattering profile. We will
show that the proposed interpretation nicely fits all the observational
evidence and is in agreement with evolutionary models of young PWN and
SNR. In particular, in Sect. 2 we present the dataset we have used,
in Sect. 3 we discuss the morphology of the X-ray halo, in Sect. 4 we
perform spatially resolved spectroscopy of the halo and its features,
in Sect. 5 we introduce the dust scattering model which explains part
of the halo emission, while in Sect. 6 we compare our findings with
evolutionary models of young PWN-SNRs.

\section{Observations}

\src\  was observed as part of the Calibration and Performance
Verification phase of the \xmm\  satellite (\citealt{jla01}). In
particular, the remnant was observed both on-axis and off-axis (for
the list of \src\ observations, see Table \ref{obs}). In this work, we
have used the array of PN CCDs (\citealt{sbd01}) and the two arrays of
MOS CCDs (\citealt{taa01}) at the focus of three X-ray mirrors of \xmm\
(\citealt{gab98}). The nominal bandwidth, angular resolution and energy
resolution of the instruments are 0.1-15 keV, $15\arcsec$ FWHM and
$E/\Delta E \sim 10$, respectively.  The original event file was screened
to eliminate the contribution of soft protons, from both flares and quiescent
emission, using the recipe given by \citet{dm04}. Since the observation
numbers 3 and 4 were found to be affected by unusually large quiescent
emission, they were discarded in the analysis of the dimmest sources
(middle and outer halo, see below).  All the analysis of XMM data was
performed with the software SAS v6.0. 

\begin{table}
\caption{Cal/PV \xmm\ Observations of \src}
\label{obs}
\medskip
\centering\begin{minipage}{8.7cm}
\begin{tabular}{lcccc} \hline
Obs. & Pointing  & \multicolumn{2}{c}{T$\rs{in}$/T$\rs{out}$ (ksec)} & Date \\
     & Location  & PN & MOS \\ \hline

OnAxis & on G21.5-0.9\footnote{At the coordinates $18^h33^m32.6^s$ and $-10^d33^m57^s$ (J2000)} & 27/18 & 30/25 & 7 Apr 2000 \\
OffAxis1 & $10^\prime$ S & 28/26 & 29/25 & 9 Apr 2000 \\
OffAxis2 & $10^\prime$ W & 28/23 & 29/27 & 11 Apr 2000 \\
OffAxis3 & $10^\prime$ N & 29/06 & 29/19 & 15 Apr 2000 \\
OffAxis4 & $10^\prime$ E & 29/06 & 29/20 & 17 Apr 2000 \\
\hline
\end{tabular}
\end{minipage}
\end{table}

\src\  was also observed as part of the calibration plan for the \chan\
satellite (\citealt{wov96}). There are $\sim 70$ observations available up to
July 2004, among which we selected only the 21 observations for which \src\
was located onto the S3 chip and at an off-axis angle less then 5 arcmin.
The observation IDs are 0159, 1433,  1554,   1717,  1769,  1771,
1839,  2873, 3693, 4353, 5166,1230, 1553, 1716, 1718, 1770, 1838, 1840,
3474,3700, 4354, and the total exposure time is 196.5 ks.  These are the
same observations used by \citet{b05}. Our set includes the set used by
\citet{shp01} plus the more recent observations that have been done (they
used 6 observations for a total of 65 ks).  Afterward, the data were
screened for bad grades and for a clean status column. All the filtered
datasets were merged together using the CIAO {\sc merge\_all} task.

\section{X-ray morphology of the halo}

\begin{figure}
  \centerline{\psfig{file=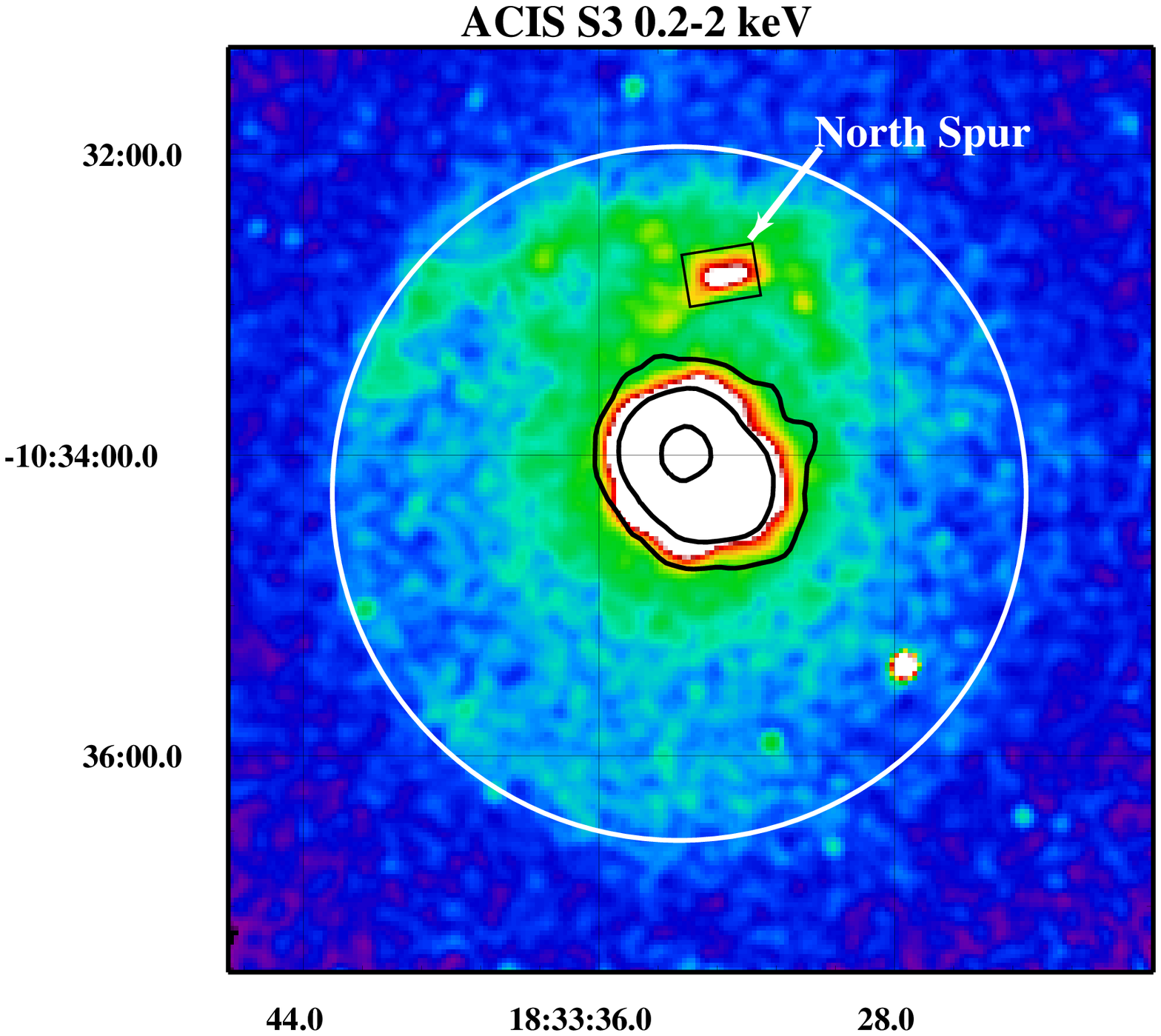,width=9.0cm}}
  \centerline{\psfig{file=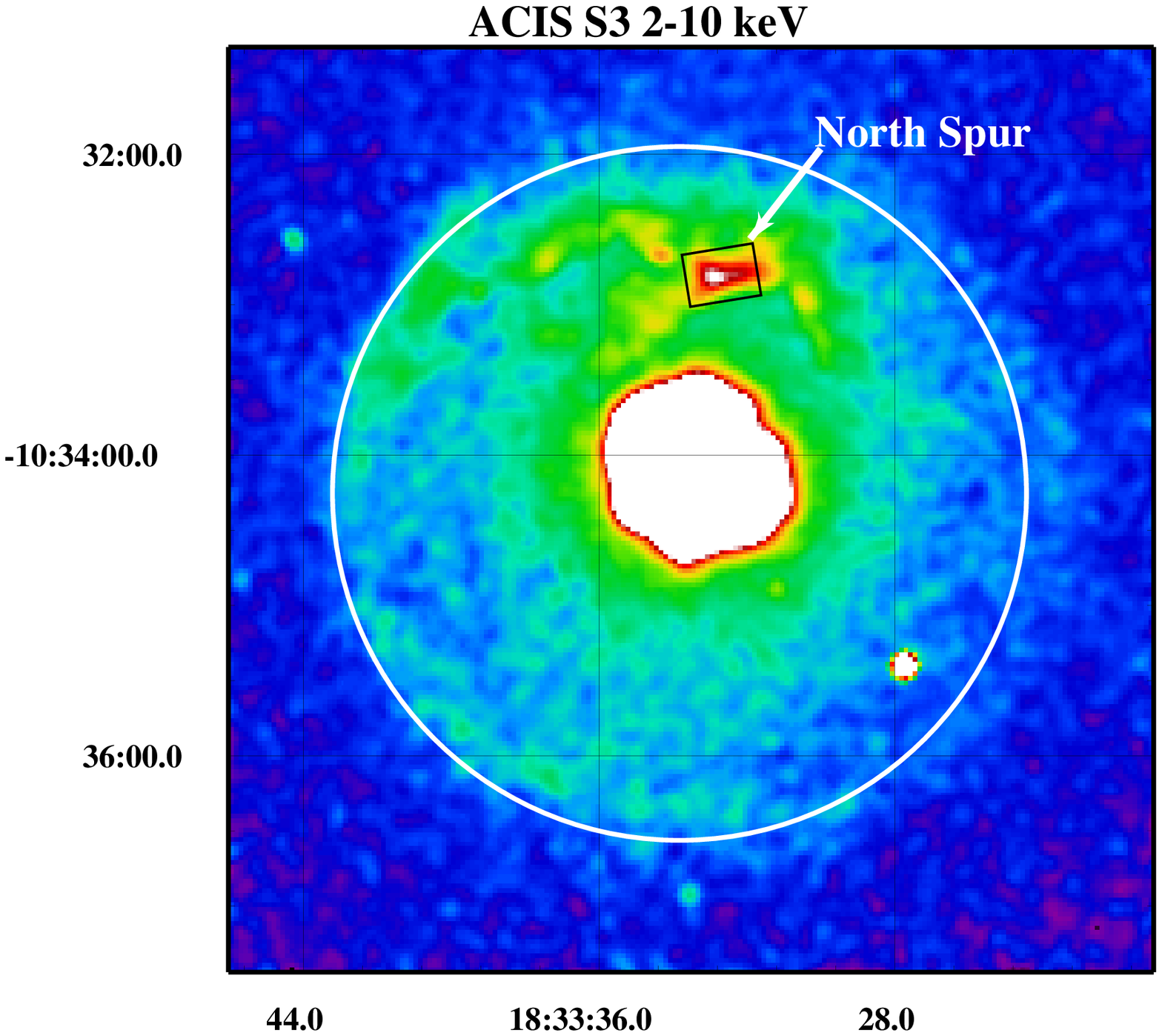,width=9.0cm}}
  \caption{\chan\ ACIS-S images of \src\  in the 0.2--2 keV ({\em
  top panel}) and in the 2--10 keV band ({\em bottom panel}). The
  color has been chosen to emphasize the weak halo emission. Black
  contours represent 22.3 GHz contours of the large-scale diffuse
  emission of \src\  at 10, 50 and 100 mJy/beam (8 arcsec HEBW) from
  \protect\citet{fhm88}. The position of the North Spur is indicated. The
  white circle fits the outer limb brightened emission of the X-ray halo,
  it has a radius of 138 arcsec ($\sim 3.3$ pc at 5 kpc) and a center
  at 8.7\arcsec\  from the peak of the PWN.}

  \label{acis}
\end{figure}

In Figure \ref{acis} we show the images of \src\  as seen by
the \chan\ ACIS-S camera in two energy bands (0.2--2.0 keV and
2.0--10.0 keV, respectively).  The figure shows the prominent
X-ray halo around the bright plerion (the latter is marked by
radio emission contours) and a compact bright feature in it at the
location $18^h33^m32.9^s$ and $-10^d32^m49^s$ (J2000), the so called
``North Spur" (NS hereafter), located at 80\arcsec\  from the center
($\sim 1.9$ pc at 5 kpc distance).
A search for counterparts of the North Spur in other wavelengths performed
both with catalog browsing (SIMBAD) and image retrieving (Skyview)
has revealed no obvious counterparts within a radius of $20\arcsec$.
The circular symmetry of the halo
at PA=210$^\circ$--315$^\circ$ is remarkable (here we adopt PA=0 at N,
positive clockwise) and a bright limb is present. The limb is outlined by
the white circle in Fig. \ref{acis}, which has a radius of 138\arcsec\
($\sim 3.3$ pc at a distance of 5 kpc) and a center located 8.7\arcsec\
(0.21 pc) in the Southeast direction with respect to the peak of the PWN.
At PA between -45$^\circ$ and 60$^\circ$ we found the NS and bright
diffuse emission apparently associated to it. This emission is in the form
of weak arc-shaped filaments which surround the North Spur and connect
it to both  the plerion and the bright limb.  In the remaining part of
the halo (approximately from PA=60\degrees and PA=180\degrees, no other
features are detected, apart from the star SS397, no limb brightening
is evident, and the halo declines more rapidly toward the background.

\begin{figure}
  \centerline{\psfig{file=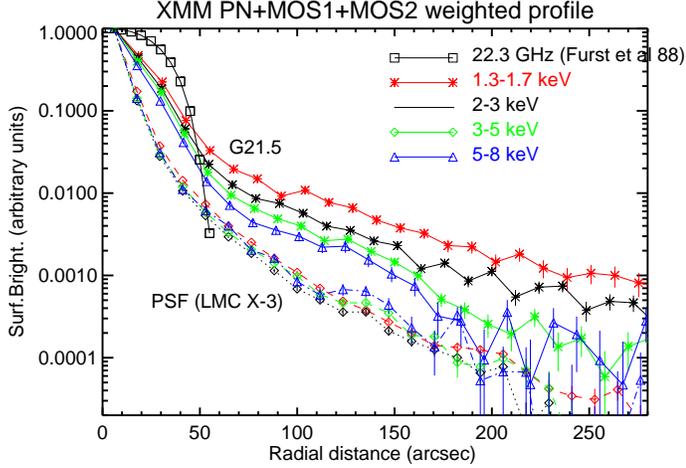,width=9.0cm}}
  \caption{X-ray surface brightness profile of \src\  as seen by \xmm. For
  comparison, we also plot the profile for a bright point source (LMC
  X-3). Moreover, we plot the profile of the source as observed by
  \protect\citet{fhm88} at 22.3 GHz.}

  \label{prof}
\end{figure}

Surface brightness profiles of the halo of \src\  have been derived
using the \xmm\  observations, and are shown in Fig. \ref{prof}. All
the profiles are computed using a weighted average of the PN, MOS1 and
MOS2 data, where the weights have been derived in each band assuming
a non-thermal spectrum with a photon index of 2 and an absorption of
$2\times 10^{22}$ cm$^{-2}$, which is a good estimate of the halo spectrum. In
order to study the profile of the ``pure'' halo, that is without the
discrete feature, we have selected only data in the PA range
60$^\circ$--160$^\circ$. All the profiles have been background subtracted
using a local background (collected in a large annulus between 280 and
320 arcsec) and vignetting corrected. Fig. \ref{prof} shows that \src\
has a sharp change in its X-ray profiles around 50\arcsec ($\sim 1.2$
pc) from the center, where the plerion profile (which is narrower than
the radio counterpart, as expected) suddenly flattens and continues out
to $\sim 250$\arcsec. 

\section{Spectral analysis}

\subsection{The halo}

\begin{figure}
  \centerline{\psfig{file=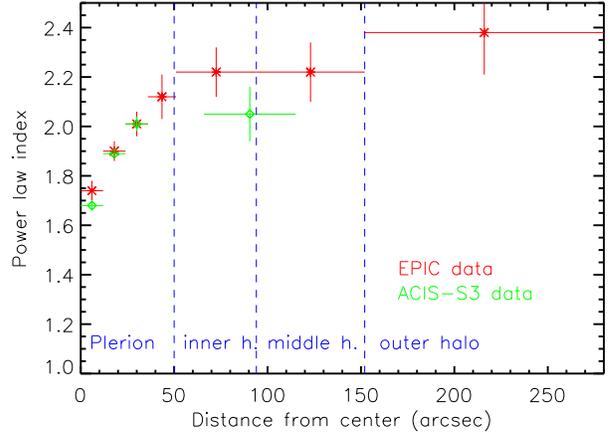,width=9.0cm}}
  \centerline{\psfig{file=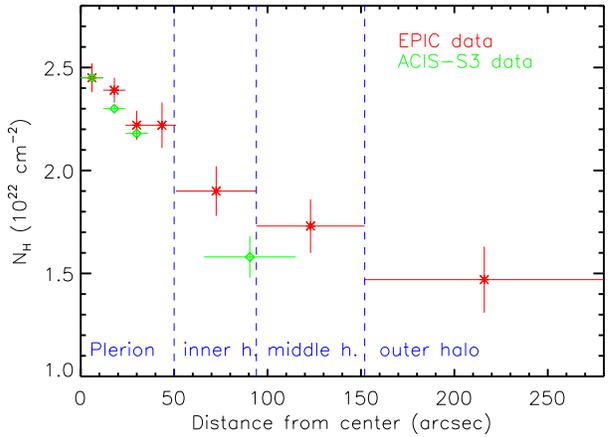,width=9.0cm}}
  \caption{Results of spectral fitting of \src\  plerion and halo data
  with a power-law model modified by interstellar absorption. We
  show the best-fit values and 90\% confidence level uncertainties for
  the absorbing column density and the photon spectral index.}

  \label{powgnh}
\end{figure}

We have performed a spectral analysis of the halo using both \chan\
and \xmm\  data.  We have generated single PN, MOS and ACIS-S3
spectra. The effective areas have been averaged, while we have used
the standard MOS response matrix for the epoch of the observations and
the on PN matrix. We have defined annulus extraction regions,
namely region 1 (0\arcsec--12\arcsec), 2 (12\arcsec--24\arcsec), 3
(24\arcsec--36\arcsec) and 4 (36\arcsec--51\arcsec) for the plerion;
5 (51\arcsec--94\arcsec, inner halo), 6 (94\arcsec--152\arcsec,
middle halo), and 7 (152\arcsec--280\arcsec, outer halo) for the halo.
Regions containing the North Spur and other filamentary structure have
been removed. We used a power-law emission model to fit the spectra.
The background has been chosen in an annulus between 302 and 330 arcsec,
excluding out-of-time events for PN.  The results of spectral fits are
shown in Fig. \ref{powgnh}.

%\begin{table}
%  \caption{List of annular extraction regions in the plerion and in the halo (best-fit values are shown in Fig. \protect\ref{powgnh}).}
%\label{results}
%\medskip
%\centering\begin{minipage}{8cm}
%\begin{tabular}{lccccc} \hline
%  Reg ($r_{in}-r_{out}$) & Norm & $\chi^2/dof$ \\
%      & ph cm$^{-2}$ s$^{-1}$ keV$^{-1}$ &  \\ \hline
%
% 1 (0\arcsec--12\arcsec) &  & 365/390\\
% 2 (12\arcsec--24\arcsec) &  & 739/395 \\
% 3 (24\arcsec--36\arcsec) &  & 338/361 \\
% 4 (36\arcsec--51\arcsec) &  & 262/253 \\
% 5 (51\arcsec--94\arcsec)\footnote{Only PA=90\degrees--220\degrees} &  & 242/205\\
% 6 (94\arcsec--152\arcsec)$^a$ &  & 181/190 \\
% 7 (152\arcsec--280\arcsec) &  & 258/261 \\
%
%\end{tabular}
%\end{minipage}
%\end{table}

The data in region 1--7 are nicely fitted with a power-law model, with
$\chi^2/dof=$ 365/390, 739/395, 338/361, 338/361, 262/253, 242/205,
181/190, 258/261, respectively. The derived spectral slope shows a
steepening in the plerion regions 1--4 which was already measured by
\citet{wbb01} and \citet{shp01}. However, in the halo regions 5--7,
no significant spectral steepening is observed and all these regions
are consistent with a photon index value of $\sim 2.3$. It is noteworthy that the
derived value of the absorbing column density is not constant in all the
regions, as would be expected. The absorption is maximum in the central
core and decreases monotonically towards the exterior parts. The difference
in $N\rs{H}$ between the center and the halo periphery is $\Delta N\rs{H} \sim
10^{22}$ cm$^{-2}$, and there is moderate evidence for flattening of
the $N\rs{H}$ decreasing trend in the halo regions.  Such a high difference
between the core and the halo can hardly be intrinsic, corresponding
to a line of sight of 3 pc of absorbing material at 1000 atom cm$^{-3}$.

\begin{figure}
  \centerline{\psfig{file=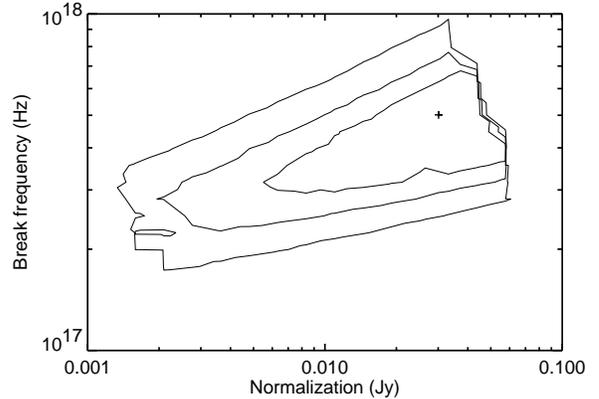,width=8.0cm}}
  \caption{68\%, 90\% and 99\% confidence level $\chi^2$ contours on the
  spectral break vs. radio flux parameters of the SRCUT model obtained
  in a fit of the X-ray spectrum of the bright limb in Fig. \protect\ref{acis}.}

  \label{srcut}
\end{figure}

Moreover, we have also selected a region to study in detail: 
the bright limb of the halo between 115\arcsec and 138\arcsec\
(PA=199\degrees--318\degrees). Spectral fits in these regions were
performed to test the presence of an additional thermal emission due to
ISM (interstellar medium) heated by the forward shock, using a combination of power-law and
the {\sc mekal} model in XSPEC v11.1 (\citealt{mgv85}) with standard
abundances. We have also tried the model ``SRCUT'' of \citet{rey98}
which represents the emission of electrons accelerated at the strong shock
of a SNR shell. The latter model was used to test if the X-ray halo
may be a non-thermal shell like SN1006 (\citealt{drb04}), G347.3-0.5
(e.g. \citealt{uat03}), RX J0852.0-4622 (e.g. \citealt{she01}), and
other young SNRs.  The thermal component is not detected in the \chan\
ACIS-S3 spectrum, and we derived an upper limit of $10^{10}$ cm$^{-5}$ to
the emission measure of 1--10 keV plasma, corresponding to an upper limit
of 0.65 cm$^{-3}$ for the post-shock density and an emitting mass $<0.045$
M$\rs\odot$\footnote{ Here and in the following we assume a distance of 5
kpc (\citealt{dsb86,bwd01}). The density and the mass scale as $d^{-1/2}$
and $d^{5/2}$, respectively. We have also assumed the swept-up mass
is located in a thin shell.}. In the \xmm\  EPIC spectrum there is a
marginal detection of an excess below 1.2 keV which requires a thermal
component with $kT=0.2-0.7$. Since this is not confirmed in the \chan\
spectrum of the same region, and since the thermal component does not
fit the spectrum in the 0.5--0.8 keV band, we do not consider it real.

As for the fit to the ``SRCUT'' model, we do not know the radio flux
at 1 GHz and slope of the radio spectrum ($\alpha$, the energy index),
since the halo has not been detected yet in radio (for a discussion on
the parameters of the SRCUT model see \citealt{drb04}). Therefore, we
constrained $\alpha$ in the range 0.3--0.6, which is typically observed in
other non-thermal shells, and we left the radio flux and spectral break
location free to vary. In this way, we may see if the extrapolation of
the X-ray spectrum back to the radio regime according to the acceleration
model is consistent with the upper limit of \citet{scs00}. The fits of
SRCUT model to the bright limb are as good as the power-law fits, and
indicate that the upper-limit, after rescaling for the different source
regions (the upper-limit is 0.08 Jy in the rim region), is $\sim 2-3$
times above the expected radio fluxes derived with the spectral fits
(see e.g. Fig. \ref{srcut}). The value obtained  for the location of
the spectral break is $2-9\times 10^{17}$ Hz, corresponding to a maximum
energy of accelerated particles of $30-85 (B/10\mu G)$ TeV, similar
to what is found in other non-thermal SNR shells.

%the spectral break is $1-5\times 10^{17}$ Hz corresponding to a maximum
%energy of accelerated particle of $24-55 (B/10\mu G)$ TeV, similar
%to what is found in other non-thermal SNR shells.

\subsection{The North Spur}

\begin{figure}
  \centerline{\psfig{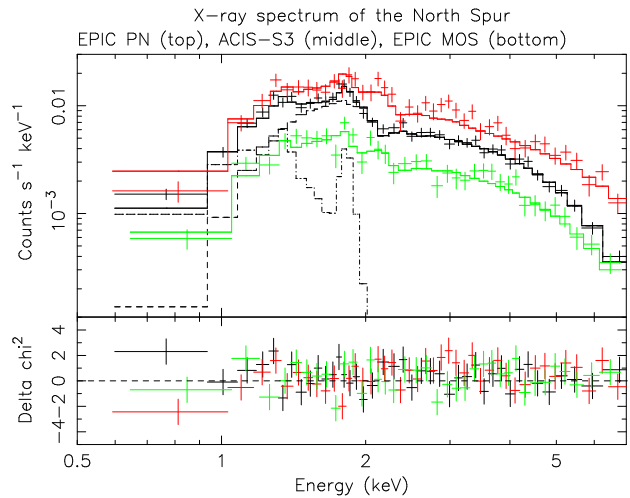}}
  \caption{\chan\  ACIS-S3 and \xmm\  PN and MOS spectra of the
  North-Spur, along with the best-fit NEI model. We also show the
  contribution of thermal and non-thermal component to the ACIS spectrum
  (dot-dashed and dashed lines, respectively).}

  \label{nspurpha}
\end{figure}

The spectrum of the North Spur was already studied by \citet{b05},
who reported the presence of an additional thermal component.
In this work, we test if the additional thermal component
is affected by
Non-Equilibrium  Ionization (NEI), which is expected to be present
in the spectra of ejecta, circumstellar and interstellar material in the SNR. 
With this aim, we
fitted the North Spur ACIS and EPIC data simultaneously with a combination
of the power-law model, the {\sc vmekal} model of the X-ray emission of
an optically thin plasma in ionization equilibrium (\citealt{mgv85}),
and the constant temperature and single ionization time NEI emission
model of \citet{blr01}. When fitting the powerlaw+vmekal and the
powerlaw+VNEI combination, we fixed the interstellar absorption value
to $N\rs{H}=2.15\times 10^{22}$ cm$^{-2}$, which is similar to the best-fit
value obtained when this parameter is left free to vary. The results
are summarized in Table \ref{rnspur}, while the ACIS and EPIC spectra
along with their NEI best-fit model are shown in Fig. \ref{nspurpha}.
The X-ray spectrum (especially the ACIS-S3 spectrum) shows signs
of the presence of two bright emission lines, namely Mg XI at 1.34
keV and Si XIII at 1.86 keV (Fig \ref{nspurpha}).

\begin{table*}
  \caption{Results obtained by a ACIS-EPIC joint spectral fit of the North-Spur X-ray emission. We fixed the interstellar absorption at $N\rs{H}=2.15\times 10^{22}$ cm$^{-2}$.}
\label{rnspur}
\medskip
\centering\begin{minipage}{17cm}
\begin{tabular}{lcccccc} \hline
  Model\footnote{PL=power-law} & $\gamma$ & Norm & $kT$ & $\tau$ & thermal flux\footnote{Unabsorbed flux in the 0.5--2.0 keV band due to the thermal component only.} & $\chi^2/dof$ \\
  $10^{22}$ cm$^{-2}$ & photon index & ph cm$^{-2}$ s$^{-1}$ keV$^{-1}$ & keV & cm$^{-3}$ s & erg cm$^{-2}$ s$^{-1}$  \\ \hline

  PL & $2.45\pm0.05$ & $2.7\pm0.2 \times 10^{-4}$ &  -  &  -  &  -  & $876.0/589$ \\
PL+VMEKAL & $2.18\pm0.04$ & $2.0\pm0.2 \times 10^{-4}$ & $0.13\pm0.06$ & - & $2.3\times 10^{-11}$ & 632.8/584 \\
PL+VNEI (M1)& $2.15\pm0.04$ & $1.8\pm0.2 \times 10^{-4}$ & $0.17(0.15-0.21)$ & $7\times 10^{11}$ & $1.5\times 10^{-11}$ & 626.7/583 \\
PL+VNEI (M2)& $2.15\pm0.04$ & $1.9\pm0.2 \times 10^{-4}$ & $0.30(0.20-0.37)$ & $1\times 10^{10}$ & $1.7\times 10^{-11}$ & 628.8/583 \\

\end{tabular}
\end{minipage}
\end{table*}

\begin{figure}
  \centerline{\psfig{file=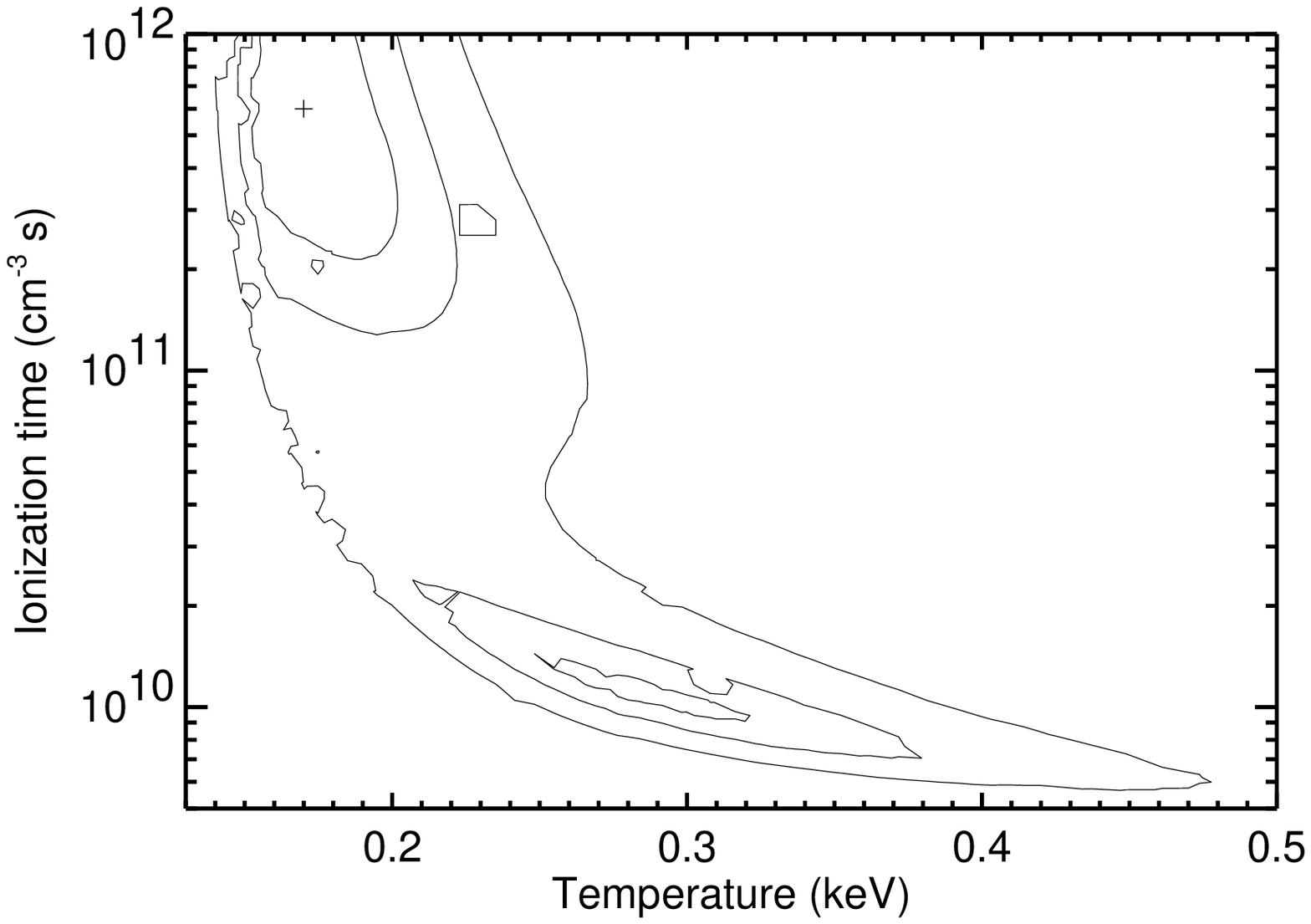,width=8.5cm}}
  \centerline{\psfig{file=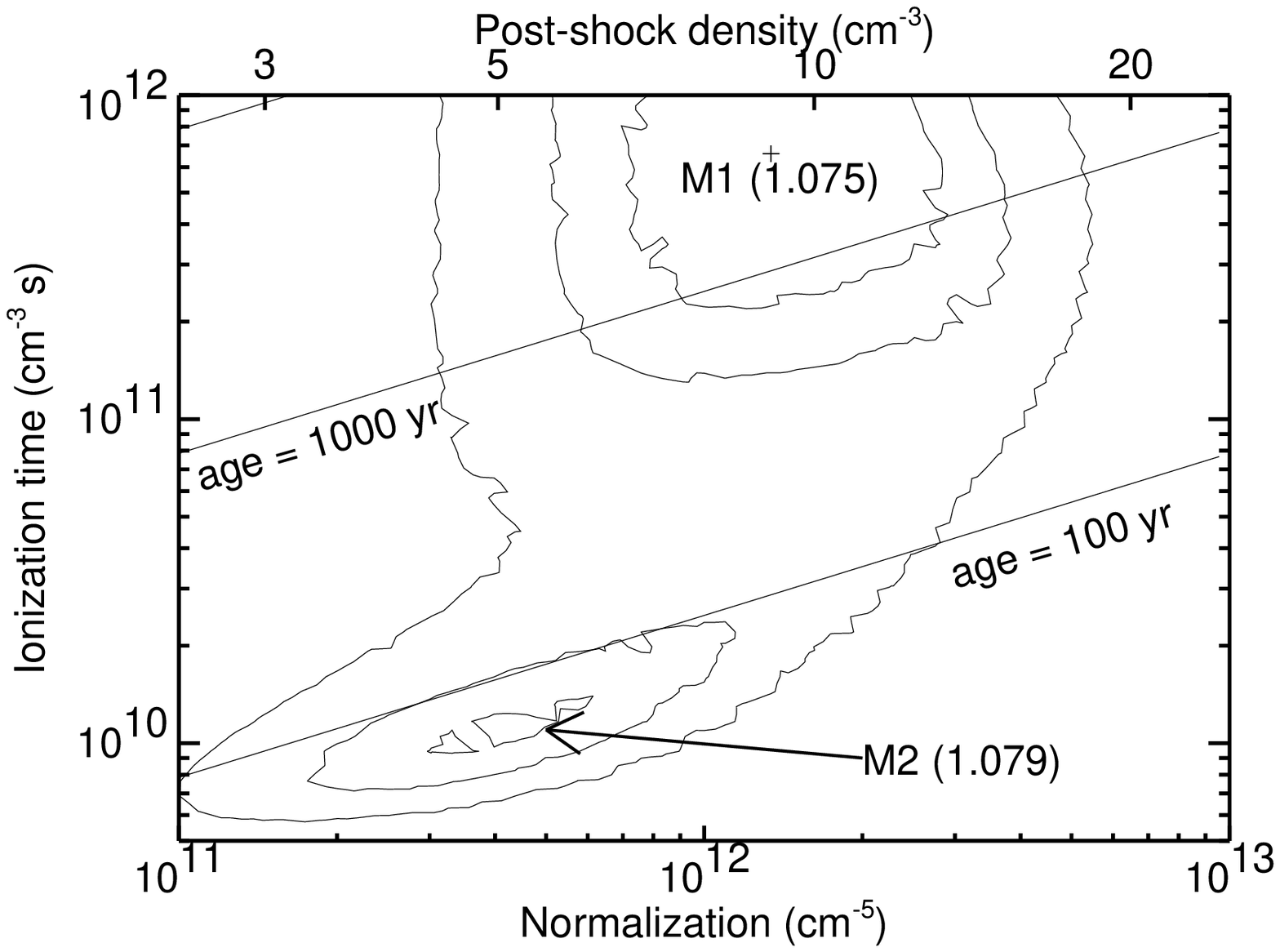,width=8.5cm}}
  \caption{Ranges of $kT$ and ionization time ({\em top panel}),
  emission measure and normalization time ({\em bottom panel}) allowed
  by the fit to North Spur data with a combination of power-law and
  NEI emission model. In the bottom panel, we also show the range of
  post-shock density and the range of ages which are compatible with the
  EM and $\tau$ values derived from the fit.}

  \label{nspur_powvnei_tauktnorm}
\end{figure}

Unfortunately, the combination of the power-law and NEI models gives two
local minima, one which represents the equilibrium situation already
found with {\sc mekal} fit ($\chi^2/dof=626.7/583$, hereafter M1),
and one which represents a plasma strongly affected by NEI conditions
($\chi^2/dof=628.8/583$, hereafter M2). Both results are presented in
Table \ref{rnspur} and displayed in Fig. \ref{nspur_powvnei_tauktnorm}.

The minimum M1, which has a slightly lower $\chi^2$ value then M2,
yields parameters similar to the ones found by \citet{b05}. On the other
hand, the minimum M2 gives an ionization time between $7\times 10^{9}$
and $2\times 10^{10}$ cm$^{-3}$ s and a temperature in the range 0.2-0.4
keV (at the 90\% confidence level), a factor of two higher than M1
(Fig. \ref{nspur_powvnei_tauktnorm}, top panel).

The NEI results at minimum M2 lead to an emitting plasma density of
$5(3-8)$ cm$^{-3}$ and mass of $0.23(0.17-0.37)$ M$_\odot$, if the
line of sight extension of the emitting plasma is equal to the chord
intersecting the sphere of the SNR shell ($\sim 6.5$ pc)
At the minimum M1 the corresponding values are $\sim 10$ cm$^{-3}$ and
$\sim 0.5$ M$_\odot$, respectively. It is noteworthy that the derived
age of the North Spur compatible with the M2 minimum is $\sim 100$ yr,
which is significantly lower than the corresponding age at M1 ($\gsim
1000$ yr, Fig. \ref{nspur_powvnei_tauktnorm}, bottom panel).

\begin{figure}
  \centerline{\psfig{file=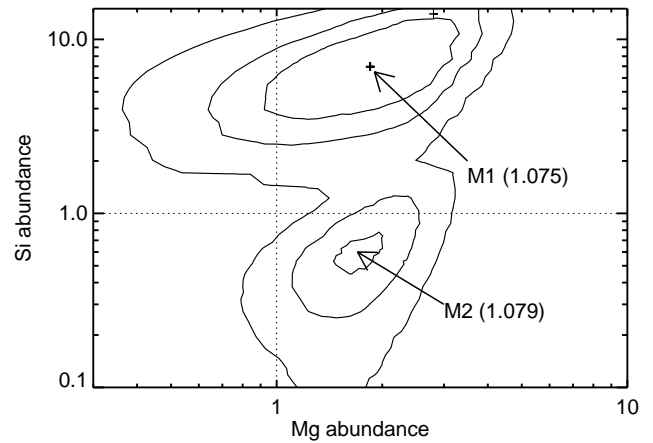,width=9cm}}
  \caption{Mg and Si abundances obtained by fitting with a powerlaw and VNEI
  model. The locations of the two minima (M1 and M2) are marked by arrows,
  and the $\chi^2$ is reported.}

  \label{nspur_mgsi}
\end{figure}

The metal abundances of Mg and Si, as measured by the minimum M1
are 0.6--3 times the solar value for Mg and 2--20 for Si, but for M2
the abundances are consistent with the solar values for both elements
(Fig. \ref{nspur_mgsi}).  The measured thermal flux corresponds to
a luminosity of $\sim 4\times 10^{34}$ erg s$^{-1}$ in the
0.5--2.0 keV band. This luminosity is a bit high compared to an ejecta knot
of Cas A, for which from the results published by \citet{lh03} we have
computed $L\rs{max} \sim 10^{34}$ erg s$^{-1}$. However, our measurement
of the unabsorbed thermal flux
is affected by large uncertainty, which can increase or decrease the
luminosity by a factor of 10. The total absorbed flux of the North Spur in the
0.5--2.0 keV band is $3.2\times 10^{-13}$ erg cm$^{-2}$ s$^{-1}$, of which $\sim
25$\% is due to the thermal component.

\section{A dust scattering model for the ``pure'' halo}

In this section, we investigate whether the diffuse emission of the \src\
X-ray halo may be due to dust scattering of  X-ray photons from the plerion.
There are reasons to suspect that dust scattering
contributes appreciably to the G21.5--0.9 X-ray profile.  One is the
large absorption column density toward this source.  Dust scattering
X-ray halos are typically found around other heavily absorbed sources
(see e.g. \citealt{ps95,sd98,nde01,vwo04}).
\citet{ps95} have shown that
there is a good correlation between the optical depth for dust scattering
and the absorption column density.  Using their empirical formula, for
G21.5--0.9 one expects $\tauone\simeq1$ (where $\tauone$ is the scattering
optical depth a 1~keV).  For the sake of illustration, $\tauone=1$ implies
a halo fractional flux of about 50\%, 20\%, 7\% and 3\%, respectively for
the energy bands 1.3--1.7~keV, 2--3~keV, 3--5~keV and 5--8~keV (these
will be the reference energy bands, used in our analysis).  However,
there is a large scatter about that correlation, so that it cannot
be taken as a safe method to evaluate $\tauone$ in individual sources.

Another effect that could be also ascribed to dust scattering is the systematic
decrease of the absorption column density with the distance from the source
center (see Fig. \ref{powgnh}), as derived from X-ray spectral analysis.
If it was a true column density variation, it would require an improbable
distribution of the foreground matter.
Instead, a lower measured absorption in the outer regions can be justified as
an artifact of the spectral analysis, in the presence of dust scattering, due
to the fact that hard X-rays are scattered at lower angles than soft X-rays.

We present here an analysis along the lines of that by
\citet{bb04}, but carried out with much better accuracy.
First, we start from data of better quality, because they are based on a longer
integration time, and with more careful selection of the directions along
which the contamination from the North Spur and the shell is the lowest.
Then, we have corrected for some small bias in the profile normalization,
which was affecting the analysis of \citet{bb04}.
Finally, we have performed a much better analysis on the parameter space.

In short, the procedure is as follows.
We assume that the dust halo is negligible in the 5--8~keV band (we
have already seen that, for $\tauone=1$, the halo fractional flux in this band
is only 3\%).
Therefore, the radial profile in the 5--8~keV band is well approximated by the
convolution of the intrinsic source profile with the instrumental Point
Spread Function (PSF).
Let us assume that the PSF is energy independent (this is a good approximation,
for EPIC onboard XMM-Newton) and also that the source
intrinsic radial profile does not change much with the energy (this assumption
will be verified ``a posteriori'').
In this case, the 5--8~keV radial profile should reproduce rather well
profiles in other bands, except for the different flux normalizations and for the further
scattering component, which becomes more prominent in softer energy bands.
For a given model of the scattering halo, the scattering component is thus
obtained as its convolution with the PSF.
Finally, the sum of the intrinsic and scattering components must be compared
with the profile measured in a given energy band (we shall use 2--3 keV).
A least $\chi^2$ analysis allows one to constrain the parameters of the
halo model.
This is a rather time-consuming procedure, because for each set of parameters
we are required to compute one convolution of 2-dimensional maps.

It is well known that, as long as the Rayleigh-Gans approximation is valid,
the scattering optical depth scales as $E^{-2}$ (where $E$ is the photon
energy), while the angular size of the halo scales with $E^{-1}$
(see e.g. \citealt{pre98} for a discussion of these scaling laws).
For the angular size, a formula similar to that of classical diffraction
applies, namely scattering angles are of the order of the ratio between
photon wavelength and dust grain size.
Using this scaling law, we can then simulate profiles for the other two energy
ranges, 1.3--1.7~keV and 3--5~keV, and compare them with observations.

However, it should be  clear that even in this sophisticated analysis we can only
check for self-consistency of a scattering halo model, while we cannot prove
for sure the dust-scattering nature of an X-ray halo.
In principle, both spatial and spectral properties of a source may conspire
to mimic shape and scaling of a dust-scattering halo.
However, the required conditions would be highly unlikely, and this is the
reason why we finally conclude that the dust-scattering halo hypothesis is confirmed
to a high confidence level.

\begin{figure}
  \centerline{\psfig{file=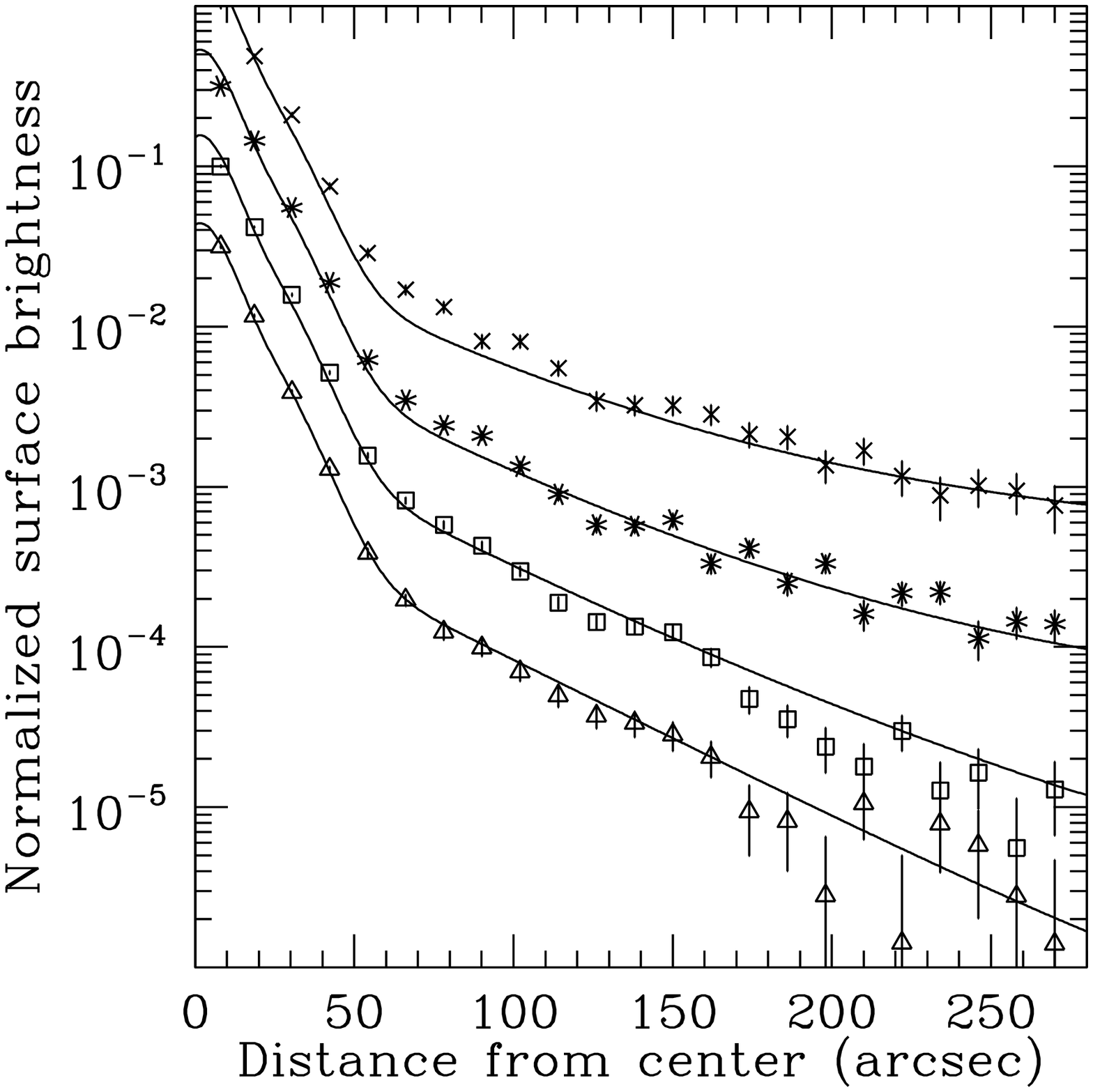,width=7.5cm}}
  \vspace{.2cm}
  \centerline{\psfig{file=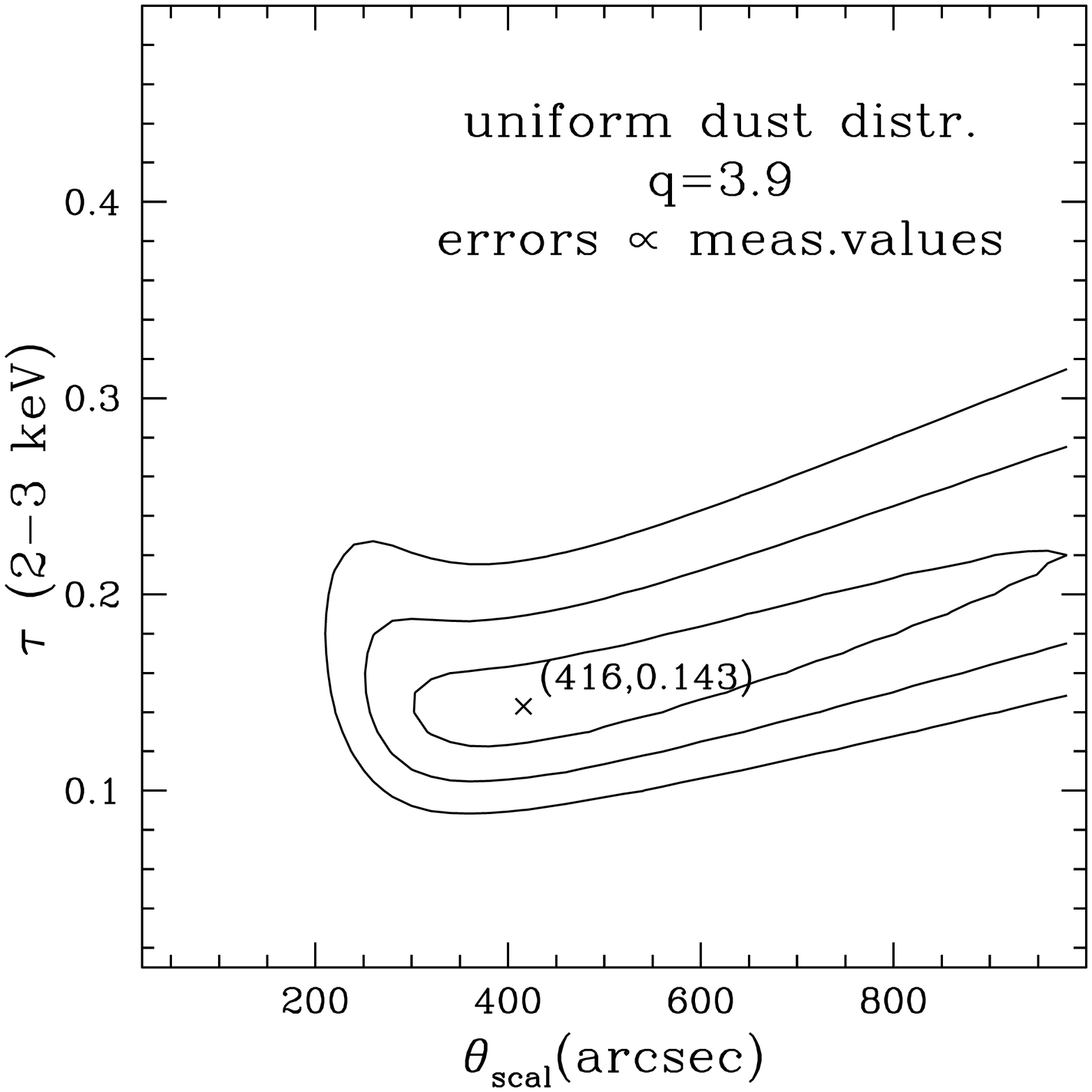,width=7.5cm}}
  \caption{{\em Top panel:} \src\  X-ray profiles with best-fit dust
  scattering model (crosses 1.3--1.7 keV, asterisks 2--3 keV, diamonds
  3--5 keV, triangles 5--8 keV). The derived dust parameters and their
  uncertainties are shown in the {\em bottom panel}.}

  \label{ds}
\end{figure}

The relevant parameters to model a dust-scattering halo are the following:
the optical depth for scattering (at the reference energy, say $\tauone$); the
dust distribution along the line of sight (here we shall use
a uniform distribution); the
size distribution of grains, usually approximated by a power-law with an upper
cutoff ($\propto a^{-q}$ for $a<\amax$; typical values are $\amax=0.17\U{\mu
m}$ and $q=3.9$, Predehl and Schmitt 1995).

For the scattering halo, we use a simplified, analytic model, that will be
described in a forthcoming paper, together with details of the procedure of
halo modeling and subtraction in individual sources.
We outline here only some basic features of this model and of halo
models in general (under Rayleigh-Gans, and single-scattering approximations).

If the space distribution of dust is uniform along the line of
sight\footnote{We have tried to relax this hypothesis by considering a
thin slab of scattering dust at a given distance. The chi-square analysis
shows that the ``uniform'' case has to be preferred. More elaborate
and complex scenarios are not justified by the present data quality.}
then the halo profile is $\propto\tht^{-1}$, for $\tht\ll\thtsc$, where

\begin{equation}
  \thtsc(x)\simeq
    \frac{3200\U{arcsec}}{(E/1\U{keV})(\amax/0.17\U{\mu m})},
\label{eq:thtcr}
\end{equation}

The range of radial distances over which we see the halo in G21.5--0.9
is about 100--300~arcsec.  Therefore, from Eq.~\ref{eq:thtcr} it follows
that we are in the $\tht\ll\thtsc$ regime.  A consequence of this (as
we have found by analyzing numerically the parameter space) is that
the best fit halo does not constrain at all the value of $q$: then,
we have assumed for this parameter the average value 3.9, as given by
Predehl and Schmitt (1995).

For the minimum $\chi^2$ fit, we have used errors proportional
to the measured values, because the radial
profile within $\sim100\U{arcsec}$ is much more uncertain than one would
infer from the (very small) statistical error (here the main uncertainty
derives from the fact that the intrinsic source profile slightly changes
with energy). The best-fit we got is shown in Fig. \ref{ds}, along
with the derived $\tau$ (in the 2--3 keV band) and $\thtsc$. We thus
estimate $\tau\rs{1keV}=5.68 \tau\rs{2-3 keV} \simeq 0.80$, with a 1$\sigma$
uncertainty of about 20\%.

\begin{figure}
  \centerline{\psfig{file=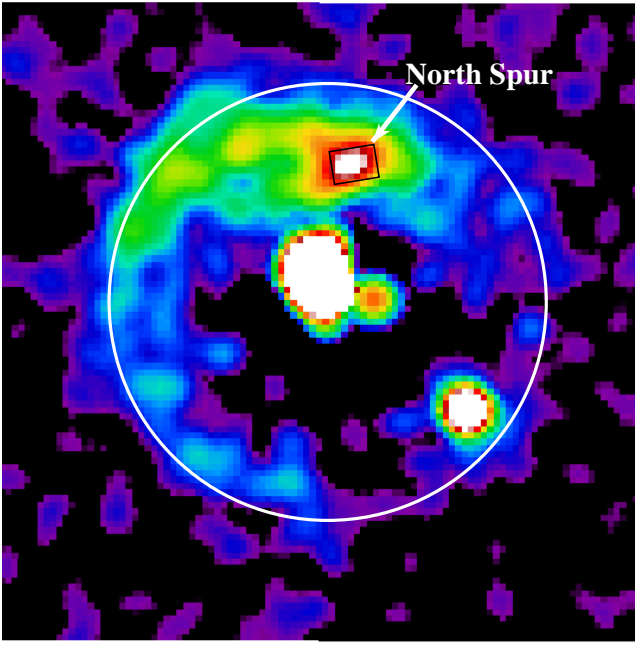,width=8.5cm}}
  \caption{Image of \src\  in the 2--8 keV with dust scattering contribution
  subtracted.}

  \label{g21nohalo}
\end{figure}

In Fig. \ref{g21nohalo} we show an image of \src\  in the band 2--8
keV with the best-fit dust scattering model profile subtracted. The
emission which remains after the halo subtraction is the residual flux
from the plerion (which is not expected to be properly modeled by
a circularly symmetric model), the North Spur, its filaments and the
shell. In particular, the shell is clearly visible from PA=180\degrees\
to 300\degrees, while it is not visible at PA=90\degrees--180\degrees.

In a separate paper, we will report more details on the dust scattering
model and the fit to the \src\  X-ray halo, including a thin slab dust
distribution and a discussion of the dust properties we found, in the
framework of general dust properties as derived from studies of X-ray
halos in other sources.

\section{A self-consistent model for the X-ray halo and its features}

\subsection{Establishing the evolutionary stage of \src}

As we have seen, the bright limb in Fig. \ref{acis} and \ref{g21nohalo} suggests
that we have detected the forward shock of \src\  expanding into the
environment of the SNR. If this is the case, we may use evolutionary models
of young SNR to infer the age of \src.  First, we argue that the remnant
is not  yet in the Sedov-Taylor (ST) phase, because the swept-up mass by
the shock front in the bright limb, as derived in Sect. 4, is very low
($<0.045$ M$_\odot$). Even taking into account the mass in the North
Spur and the other filaments not analyzed, the total swept-up mass in
the halo is less than 0.5 M$_\odot$.

\citet{rac05} has discussed the interaction of a young core-collapse
SNR with its environment, taking into account different supernova types,
namely SN1987A-like class, IIP, IIL/b, and Ib/Ic. Because of the low mass
observed in the halo, we may safely discard a dense environment for
\src\  like the RSG wind usually found for Type IIL/b. In fact, it can
be shown that for an observed radius of 3.3 pc (Sect. 3), the remnant
of a Type IIL/b with a dense RSG wind should have swept-up most of the
total ejecta mass and a few solar masses of circumstellar medium.

\begin{figure}
  \centerline{\psfig{file=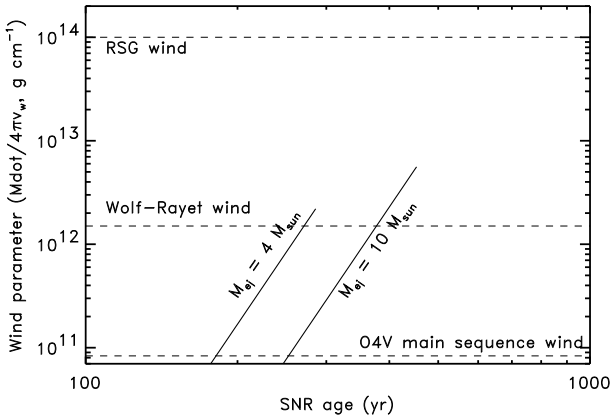,width=8.5cm}}
  \caption{The CSM interaction model of \protect\citet{rac05} adapted
  to the \src\  case. The two segments are the loci of allowed wind
  parameter and SNR age values calculated for 4 and 10 M$_\odot$ ejecta
  masses and an upper-limit of 0.1  M$_\odot$ for the CSM swept-up in
  a shell, in agreement with the results obtained in Sect. 4.2 for
  the bright halo. The wind parameters for three kinds of stars are also
  shown, a RSG wind (which turns out to be incompatible with the data
  for any reasonable value of the system parameters), a WR star wind
  (one of the viable interpretations for the \src\  progenitor), and a
  O4V normal star, shown only as a reference for the minimum value the
  wind parameter may have.}

  \label{cs}
\end{figure}

Type IIP and Ib/Ic are interesting possibilities. For Type IIP,
\citet{rac05} notes that the low mass loss during the RSG phase would
result in a small region ($r<1$ pc) of dense wind surrounded by a more
diffuse extended bubble created during the main sequence phase. The
interaction with the resulting double layer wind structure has not been
modeled in detail but swept-up masses are of the order of 0.1 M$_\odot$,
so in agreement with observations. For Type Ib/Ic, it is expected that
they undergo a Wolf-Rayet star phase with high mass-loss and fast wind which
sweeps the earlier RSG bubble in a shell at several parsecs from the
center. The evolution of the SNR in this medium may also be complicated
and requires numerical simulation (\citealt{dwa01}). However, if we assume
that the star has been a WR object long enough to produce an extended wind
component, we may use the interaction model developed by \citet{rac05}
for the circumstellar interaction of an RSG wind, but with the parameters
appropriate for a WR wind ($E_{51}=1$, $M\rs{ej}=4$ M$_\odot$, $\dot M =
3\times 10^5$ M$_\odot$ yr$^{-1}$, wind velocity $v\rs{w}=1000$ km s$^{-1}$,
$D=\dot M/4\pi v\rs{w}=1.5\times 10^{12}$ g cm$^{-1}$). The model assumes that
the progenitor star had a radiative envelope and treats the interaction in
the thin shell approximation.  Although the progenitor star of G21.5--0.9
may not have had a radiative envelope, the steep outer power law with a
relatively flat central density distribution is probably a reasonable
approximation to the density profile.  We assume that the interaction
shell is still in the outer steep power law part of the supernova density
profile; this assumption can be verified for the parameters we find for
G21.5--0.9. In this case, we derive a CSM swept-up mass of 0.1 M$_\odot$,
an ejecta swept-up mass of 0.2 M$_\odot$, which are again roughly in
agreement with observed mass in the North Spur and the upper-limit of the
shell. The derived shell velocity is $\sim 7000$ km s$^{-1}$, and the
age is $\sim 290$ yr.  For any reasonable set of wind parameters, the upper-limit
on the swept-up mass derived in Sect. 4.1 implies an age lower then
$\sim 450$ yr. This is shown in Fig. \ref{cs} where allowed SNR ages are
plotted for different wind parameters and ejecta masses according to the
self-similar solution of the CSM interaction model of \citet{rac05},
adapted to the observed size of \src, and considering only a swept-up
CSM mass below 0.1 M$_\odot$, in agreement with the observations.

Therefore, the spectral results for the shell and the North Spur,
when combined with reasonable scenarios of the SN environment, seem
to indicate in any case an early evolutionary stage for the SNR and
a progenitor without extended dense winds. We may now see if this is
also compatible with the size of PWN and some appropriate model for
PWN-SNR interaction.  Several authors (\citealt{rc84,cf92,vag01}) give
analytical expressions which relate the radius of the PWN shock and the
SNR blastwave. We have applied these equations to the case of \src\
using a SNR shock radius of 3.3 parsec and a PWN shock radius of 1.2
parsec, under the assumption of a uniform ejecta density and constant
spin-down luminosity.  We typically get $E\rs{pwn}=L_0t \sim 10^{49}$
erg and an age of $\sim 500$ yr from these models.  While the age is in
agreement with the CSM interaction model, the corresponding (constant)
spin-down luminosity is very high especially considering that the X-ray
luminosity is considerably less than that observed from the brightest
PWNe, so this seems to suggest that significant spin-down has occurred
at the early stage of the PWN evolution.

\citet{rac05} has developed a model of PWN interaction with SN ejecta,
considering power-law density profile for the ejecta and the pulsar
luminosity decay. In the framework of his model, it is convenient to
compare the PWN internal energy with the pulsar luminosity and the shell
kinetic energy.  The equipartition energy of the nebula can be found in
the following way

\begin{equation}
  E\rs{min} = 3.2\times 10^{47} \left(\frac{L\rs{b}}{3.6 \rm{Jy}}\right)^{\frac{4}{5}}
  \left(\frac{R\rs{PWN}}{1.2\rm{pc}}\right)^{\frac{9}{7}}
  \left(\frac{\nu\rs{b}}{540\rm{GHz}}\right)^{\frac{2}{7}}
\end{equation}

where $L\rs{b}$ is the spectral luminosity at the break frequency $\nu\rs{b}$
(values for \src\  are from \citealt{bnc01}), the particle energy indexes
before and after the break are $p_1=1$ and $p_2=3$, respectively (adapted
from \citealt{rac05}). The nebula internal energy is usually within a
factor of a few of $E\rs{min}$, say $E\rs{int}=10^{48}$ erg. \citet{shp01}
found $\dot{E}=3\times 10^{37}$ erg s$^{-1}$, based on the $\dot{E}-L\rs{X}$
relation, implying $\dot{E}t=4.7\times 10^{47}t\rs{500}$ erg, where
$t\rs{500}$ is the age in units of 500 yr.  The kinetic energy in the PWN
is $\sim 10^{49}$ erg, which is determined by the supernova model and
the age.  This set of energies is consistent with the model if the ratio
of the age and the initial spin-down timescale $t/\tau$ is in the range
1--10 (see Fig. 1 in \citealt{rac05}), thus suggesting that significant
spindown can occur at the early stage of the PWN evolution.

\begin{table}
  \caption{A comparison of \src\  linear size and equipartition energy with a set of similar plerion with known age. From the comparison, one would guess a \src\  age in the range 800--1600 yr.}
\label{comp}
\medskip
\centering\begin{minipage}{9cm}
\begin{tabular}{lccccc} \hline\hline
  SNR & 0540-69 & Kes75 & G11.2 & G54.1 & \src \\ \hline
  R (pc) & 0.9 & 1.4 & 0.9 & 1.2 &  1.2 \\
  Age    & 800 & 1000 & 1600 & 1500 &  ? \\
  $E\rs{min}$ & $5\times10^{47}$ & $10^{48}$ & $3\times10^{46}$ & $8\times10^{46}$ & $4\times10^{47}$ \\ \hline

\end{tabular}
\end{minipage}
\end{table}

Finally, we should note that the comparison of the equipartition energy
and linear dimension of \src\  with other PWNe, summarized in Table
\ref{comp}, seems to suggest an age of 800--1600 yr, older then what we
found on the basis of the circumstellar interaction model.  One way of
having the PWN expand more rapidly is to have a lower density supernova,
but this generally requires a lower mass or higher energy, which also
reduce the age from the circumstellar interaction model.

\subsection{On the nature of ``North Spur''}

In this section, we review a physical interpretation for the North
Spur which is compatible with the observational results. Fig. \ref{acis}
shows that the spur is located at 35\arcsec\  (0.8 pc) outside the
plerion rim, with a low surface brightness in between. This is at odds
with an interpretation in terms of material swept-up by the plerion,
because in that case we expect the thermal emission adjacent or inside
the plerion non-thermal emission (as in, e.g., 3C58, \citealt{bwm01}).

Therefore, a possible valid interpretation may be that the North Spur is
an ejecta clump hit by the reverse shock. In this case, the position of the
forward shock ($R\rs{fs}$), as traced by the bright limb in the halo, and
the contact discontinuity ($R\rs{c}$), as traced by the emission from the
North Spur, can be compared with self-similar solutions of interaction
of ejecta with an external medium as worked out by \citet{che82a}. For
a model in which ejecta with steep power-law outer density distribution
($n>5$, where $n$ is the power-law index) are interacting with the free
stellar wind of the massive progenitor ($s=2$, where $s$ is the power-law
index of stationary medium, usually mass-loss from progenitor star),
the forward shock is expected to be at $\lsim 1.3-1.4R\rs{c}$, unless
it has already entered the flat part of ejecta density distribution.
Unfortunately, because of projection effects, the exact location of the
North Spur inside the remnant is not exactly known, but it is between
$1.0R\rs{c}$ and $1.7R\rs{c}$, so in general agreement with expectation.
As already noted in Sect. 4.2, the density, mass and luminosity estimates
for the North Spur are not unusual for ejecta in young SNRs. However,
the measured temperature is lower then the one expected for emitting
X-ray ejecta and usually observed in other young SNRs originating from
a Type II SN (e.g. Cas A, \citealt{gkr01}), thus casting some doubt on
this interpretation.

A different explanation for the North Spur which takes into account
the low X-ray temperature may be the following.  We have seen the a
progenitor SN of Type IIP is in agreement with the observationally
derived masses. \citet{rac05} show that in case of a Type IIP SN, the
circumstellar interaction should occur early (radius $< 1$ pc), followed
by adiabatic expansion of the reverse shocked ejecta.  This adiabatic
expansion could give rise to the relatively cool emission which is
observed.  The mass is not expected to be high and it is in agreement
with the observed value of $\sim 0.2$ $M_\odot$.  Other implications of
this scenario are a young age and near solar abundances (the envelope
of the the Type IIP SN), which are both in agreement with the age and
abundances derived at the minimum M2 (Fig. \ref{nspur_powvnei_tauktnorm}),
and the dynamical age of the shell derived in the previous section.

We also note that the data indicate that the North Spur has an {\em
intrinsic} non-thermal tail in the spectrum that is modeled with a
power-law ($\gamma=2.15$, Table \ref{rnspur}). One explanation for
the non-thermal emission is synchrotron radiation from particles accelerated
in a shock. 

% In case of the ejecta interpretation this explanation has
% lot of difficulties. In fact, \citet{edb05} has showed that particle
% acceleration in expanding ejecta is orders of magnitude less efficient
% that at main shock without magnetic field amplification. Even with
% amplification, electron are not accelerated above $mc^2$, and therefore
% they are not expected to emit X-ray photons by synchrotron.

\section{Conclusions}

We presented an extensive analysis of \chan\  and \xmm\  X-ray data
of the radio-quiet halo around the plerion \src. We included in our
analysis all the public observations of the source available up to
now. We showed that the halo morphology is composed of two components:
diffuse emission and some bright knots and filaments. We confirmed
the detection of X-ray thermal emission in the brightest knot, the
``North Spur'', already detected by \citet{b05}, and we presented a more
detailed spectral analysis of this object which included Non-Equilibrium
Ionization, and which yielded a very young age (100--1000 yr) and
abundances compatible with solar values. We have presented various
interpretations for the origin of this knot, and we argue that it can
be due to ejecta interaction with the H envelope of a Type IIP SN.

We have also detected a bright limb in the east part of the halo, located
at 3.3 pc from the center. The limb is dominated by non-thermal X-ray
emission, probably due to particle acceleration at the fast forward
shock, while the upper limits for the post-shock density and emitting
mass are 0.65 cm$^{-3}$ and 0.045 M$_\odot$ in this region.

We showed that the diffuse emission from the halo is due to dust
scattering of X-rays from the plerion, and we explored which range 
of dust parameters is compatible with observations.

We argued that our data are inconsistent with an explanation of the halo
in terms of a plerion extension, as suggested by previous works. We
exclude the possibility that the system is in Sedov-Taylor stage on
the grounds that the swept-up mass is very low.  Moreover, by applying
a model of CSM interaction of a young SNR shock wave to the bright limb
data, we argue that \src\ is in a very early evolutionary stage, with an
age between 250 and 500 yr. A comparison of the plerion linear size and
equipartition energy with the corresponding values of plerions with known
age suggest a longer age (800-1600), while the comparison with a model
of SNR-PWN interaction for young remnant allows age as low as 500 yr and
indicates that spindown has already occurred. Putting together all the
estimates, it seems that a very reasonable range for the age of \src\
is 200--1000 yr. The lack of an historical supernova associated with
this object is not a surprise given the high extinction in this direction.

If our conclusions are correct, the PWN may be expanding at a few times
1000 km s$^{-1}$, and a comparison of X-ray archive images at a baseline
of 5 yr may lead to the direct detection of PWN expansion. Moreover,
additional X-ray observations would be required to study in more detail the
thermal emission both from the forward shock (if any) and in the North Spur.
Deeper radio observations of the North Spur and the bright limb which
lead to detection of these objects would shed light on the nature of
their non-thermal emission. The required radio sensitivity at 1 GHz
to detect the limb is 2--3 times below the current upper-limit. As for
the halo X-ray diffuse extended component, if it is indeed due to dust
scattering, it should not have any radio counterpart. The detection of
the pulsar would be of great value to further constrain the evolutionary
stage of the system.

\vspace{1cm}

{\em Note added after acceptance:} In April 2005, during the referee
review of this manuscript, a paper by \citet{ms05} appeared, reporting the
results of a spectral analysis based on a {\em Chandra} dataset larger
then the one used by us. The results of \citet{ms05} are consistent
with our ones, in particular the trend in the interstellar absorption
and power-law index, and the presence of limb brightening in the east
quadrant.

\begin{acknowledgements}

RAC was partially supported by NASA grant NAG5-13272.  We are grateful for
the stimulating atmosphere and support provided by the ISSI (International
Space Science Institute, Bern) workshop on the ``Physics of Supernova
Remnants in the XMM-Newton, Chandra and INTEGRAL Era.''.  EvdS was
partially supported by PPARC. FB and RB were partially supported by INAF
grant PRIN2003 ``The impact of high-resolution X-ray data on the study
of SNRs''.

\end{acknowledgements}

\bibliographystyle{aa}
\bibliography{references}

\end{document}